\documentclass[twocolumn]{aastex63}
\submitjournal{AJ}

\shorttitle{Study of star clusters in M83 with a CNN}
\shortauthors{Bialopetravi\v{c}ius and Narbutis}

\begin{document}

\title{Study of star clusters in the M83 galaxy with a convolutional neural network}

\correspondingauthor{Jonas Bialopetravi\v{c}ius}
\email{jonas.bialopetravicius@gmail.com}
\email{jonas.bialopetravicius@ff.vu.lt}

\author{Jonas Bialopetravi\v{c}ius}
\affiliation{Astronomical Observatory, Vilnius University, Saul\.{e}tekio av. 3, LT-10257 Vilnius, Lithuania}

\author{Donatas Narbutis}
\affiliation{Astronomical Observatory, Vilnius University, Saul\.{e}tekio av. 3, LT-10257 Vilnius, Lithuania}
\affiliation{Center for Physical Sciences and Technology, Saul\.{e}tekio av. 3, LT-10257 Vilnius, Lithuania}

\begin{abstract}
We present a study of evolutionary and structural parameters of star cluster candidates in the spiral galaxy M83. For this we use a convolutional neural network trained on mock clusters and capable of fast identification and localization of star clusters, as well as inference of their parameters from multi-band images. We use this pipeline to detect 3,380 cluster candidates in Hubble Space Telescope observations. The sample of cluster candidates shows an age gradient across the galaxy's spiral arms, which is in good agreement with predictions of the density wave theory and other studies. As measured from the dust lanes of the spiral arms, the younger population of cluster candidates peaks at the distance of $\sim$0.4 kpc while the older candidates are more dispersed, but shifted towards $\gtrsim$0.7 kpc in the leading part of the spiral arms. We find high extinction cluster candidates positioned in the trailing part of the spiral arms, close to the dust lanes. We also find a large number of dense older clusters near the center of the galaxy and a slight increase of the typical cluster size further from the center.
\end{abstract}


\section{Introduction}

Star clusters are gravitationally bound groups of stars that form in molecular clouds. \cite{2003ARA&A..41...57L} stated that the vast majority of all stars are formed as parts of clusters. However, \cite{2010MNRAS.409L..54B} found that only a small fraction of stars form in clusters where they are influenced by their surroundings. \cite{2020SSRv..216...69A} compiled a recent general review of cluster populations at  various evolutionary stages and in diverse galactic environmental conditions for nearby galaxies. Star clusters are unique tracers of their host galactic properties and therefore an invaluable subject of study for galaxy dynamics and evolution. Furthermore, there is a pressing need to extend extragalactic cluster catalogs to lower masses, because they tend to be under-represented in star cluster observations of distant galaxies \citep{2019ARA&A..57..227K}. There have been numerous studies on star cluster detection and parameter inference. Prominent examples include Stochastically Lighting Up Galaxies (SLUG) \citep{2015MNRAS.452.1447K}, which is one of the most mature codes in stochastic cluster population simulation and inference, and The Panchromatic Hubble Andromeda Treasury (PHAT) survey \citep{2012ApJS..200...18D}, which is one of the largest homogeneous star cluster studies based on the Hubble Space Telescope (HST) data.

\begin{figure*}
    \centering
    \includegraphics[width=1.0\textwidth]{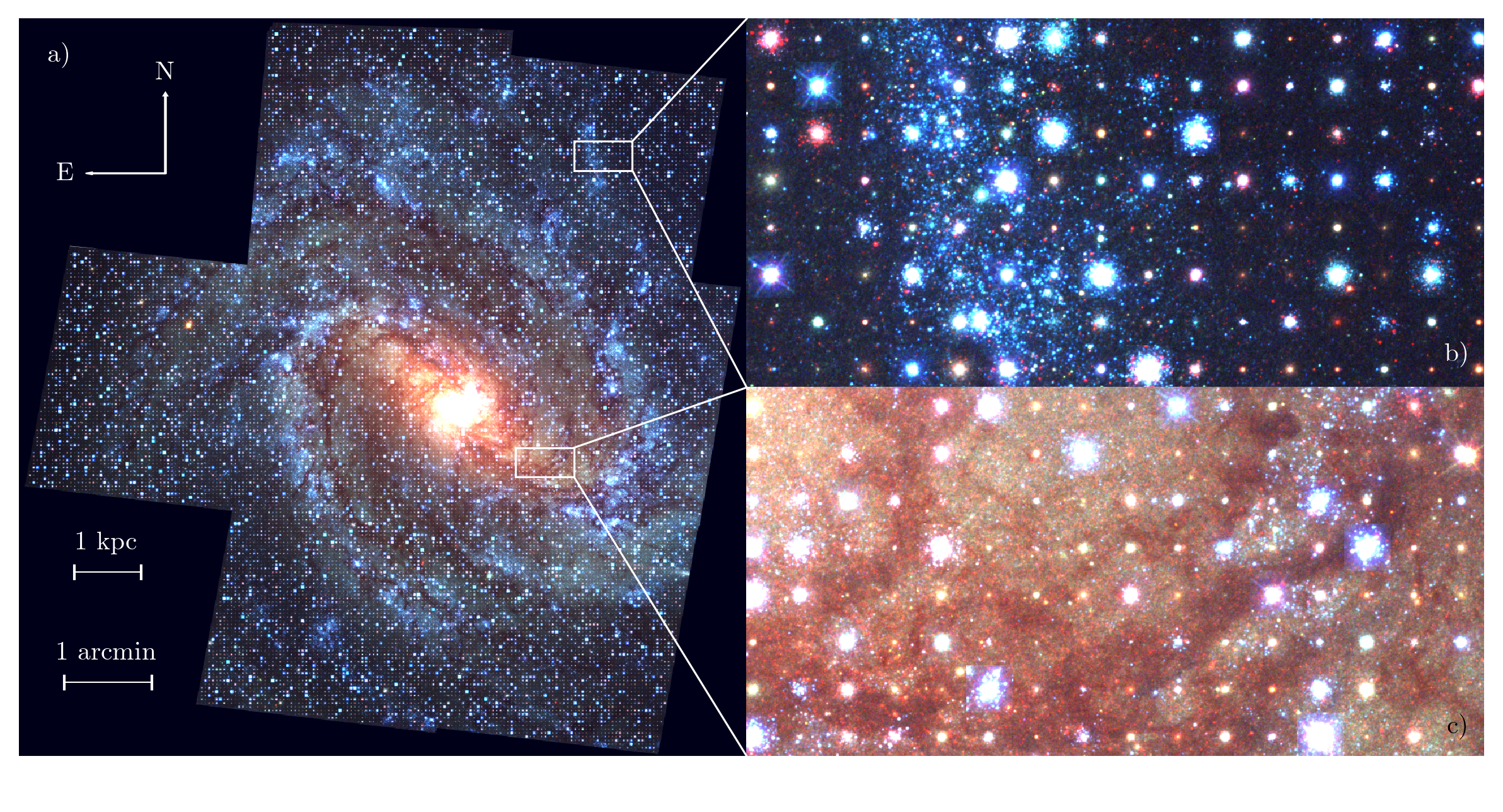}
    \caption{Panel a shows the mosaic of seven HST WFC3 fields of the M83 galaxy in three photometric passbands: F336W (blue), F438W (green), and F814W (red). Mock clusters used for artificial cluster tests ($\sim$26,000 in total) are superimposed on the image in a dense grid, placing a cluster every 64$\times$64 pixels. Panels b and c show zoomed-in areas of the galaxy's outskirts and central region respectively. The superimposed mock clusters are of various luminosities and appearances due to the random sampling of their ages, masses, extinctions, and sizes.}
    \label{fig:mocks_on_galaxy}
\end{figure*}

There is an accelerating uptake of convolutional neural networks (CNNs) in the field of astronomy. Recent examples include astrophysical source search and classification \citep{2019Galax...8....3L,2020ApJS..248...20H}, galaxy morphology exploration \citep{2020ApJ...895..112G}, spectra analysis \citep{2020PASP..132d4503Z}, supernova \citep{2020ASPC..522..451L} and contaminant \citep{2019ASPC..523...99P} detection among many others. This trend opened the door for the analysis of the exponentially increasing amount of sky survey data. In \citet[][Paper I]{PaperI} we applied a CNN to infer age, mass, and size of M31 star clusters. This was followed by \citet[][Paper II]{PaperII}, where we extended the CNN to additionally infer extinction, as well as to provide metrics describing the star cluster detection certainty, inspired by ``Telescopes don't make catalogues!'' \citep{2010EAS....45..351H}.

This work focuses on the application of a CNN to star cluster candidate detection and parameter inference on the M83 galaxy HST data. A full pipeline was constructed that performs star cluster analysis starting from multi-band images and resulting in evolutionary, structural, and environmental parameter estimates for 3,380 cluster candidates. This sample was used to analyse the dependence of cluster parameters on their galactocentric distance and positions relative to the spiral arms.

The paper is structured as follows. In Sec. \ref{sec:data} we present the HST image data and the mock clusters used in the training and testing of the CNN. In Sec. \ref{sec:method} we present how the CNN is applied to cluster search, as well as the artificial cluster tests used to measure its performance. Finally, in Sec. \ref{sec:results} we explore the results on the detected sample of cluster candidates through various astrophysical aspects, and in Sec. \ref{sec:discussion} we discuss the limitations of the method and some of the caveats of our results.

\hfill \break

\hfill \break

\section{Data} \label{sec:data}

\subsection{M83 mosaics}

We used M83 images observed with the HST Wide Field Camera 3 (WFC3), which were obtained from the Mikulski Archive for Space Telescopes\footnote{https://doi.org/10.17909/T96888}. Drizzled mosaics of seven WFC3 fields were used, which are photometrically and astrometrically aligned; the details of the mosaic processing are provided by \cite{2014ApJ...788...55B}.

Although the HST survey was performed in numerous passbands, we selected F336W, F438W, and F814W mosaics, which cover the whole galaxy without gaps and allow homogeneous cluster search and parameter inference throughout the galaxy. Overall, the mozaics for these three passbands are of very good quality, with minimal artefacts and relatively uniform depth; exposure times for each passband are in the following ranges: F336W 1900--2600 s, F438W 1900--2900 s, F814W 1200--2700 s.

\subsection{Mock clusters} \label{sec:mock_clusters}

\begin{deluxetable*}{rrcl}
    \tabletypesize{\scriptsize}
    \tablecaption{Star cluster parameters}
    \tablehead{\colhead{Parameter} & \colhead{Notation} & \colhead{Value} & \colhead{Comment}}
    \startdata
    \multicolumn{3}{c}{\it Sampled and inferred} & \\
    Age & $\log(t/{\rm yr})$ & [6.6, 10.1] & discrete, step of 0.05 dex\tablenotemark{a} \\
    Mass & $\log(M/{\rm M_\odot})$ & [3.5, 5.5] & continuous \\
    Extinction & $A_V$ [mag] & [0, 3] & discrete, step of 0.1 mag\tablenotemark{a} \\
    Size & $\log(r_h/{\rm arcsec})$ & [$-1.4$, $-0.4$] & continuous EFF parameter sampling \\
    \hline
    \multicolumn{3}{c}{\it Fixed} & \\
    Metallicity & $Z$ & 0.03 & constant \citep{2019ApJ...872..116H} \\
    Distance & $d$ [Mpc] & 4.5 & constant \citep{2003ApJ...590..256T} \\
    \hline
    \multicolumn{3}{c}{\it Inferred} & \\
    Visibility & $visibility$ & ($\sim$0.1, $\sim$1,000) & continuous, a proxy for signal-to-noise \\
    Cluster or background & $class_{c/b}$ & [0, 1] & continuous, detection certainty \\
    \hline
    \multicolumn{3}{c}{\it Computed} & \\
    Density & $\log(\rho_h/{\rm (M_\odot \cdot pc^{-3})})$ & [$\sim$0, $\sim$5] & continuous, computed within $r_h$ \\
    \enddata
    \tablenotetext{a}{From Padova isochrone bank with standard extinction law ($R_V=3.1$).}
    \label{tbl:parameter_ranges}
\end{deluxetable*}

Fig. \ref{fig:mocks_on_galaxy} shows the M83 mosaic comprised of three passbands (F336W, F438W, and F814W) used in this study, with superimposed mock clusters. For the training of a CNN a large number of mock cluster images with corresponding cluster parameters are required. Mock clusters are also needed to perform artificial cluster tests, which allow us to estimate the network's performance in an object search scenario. Our clusters were generated with four main randomly sampled parameters: age (defined by an isochrone), mass (initial at cluster's birth), extinction (assumed to be the same for all stars), and size (characterized by $r_h$ -- the radius of a circle on the sky enclosing half of the stars of a cluster).

We generated mock clusters in three stages. First, cluster age, mass, extinction, and size were sampled from pre-set ranges. The general procedure for this is described in Paper II, while the ranges are summarized in Table \ref{tbl:parameter_ranges}. The parameters are not sampled on a grid, but rather uniformly and independently of each other. Secondly, stars of the cluster were generated by Monte-Carlo sampling their masses from an initial mass function (IMF) \cite{2001MNRAS.322..231K}. Finally, the cluster's stars were drawn in an image by: 1) converting star absolute magnitudes to photon counts per second using Padova isochrones\footnote{http://stev.oapd.inaf.it/cgi-bin/cmd} and assuming the distance of 4.5 Mpc \citep{2003ApJ...590..256T}, 2) generating star positions according to the Elson-Fall-Freeman (EFF) profile \citep{1987ApJ...323...54E}, 3) rendering the stars using TinyTim\footnote{http://tinytim.stsci.edu/cgi-bin/tinytimweb.cgi} PSFs. Only clusters of $m_{F336W}<24$ mag, $m_{F438W}<23.5$ mag, and $m_{F814W}<23$ mag were selected for further use. See Figs. 1--4 in Paper II for visualizations of mock clusters, as well as their parameter values.

\section{Method} \label{sec:method}

\subsection{Convolutional neural network}

We used a CNN based on the ResNet50 architecture \citep{2015arXiv151203385H}, detailed in Paper II, capable of simultaneously inferring age, mass, extinction, size, $class_{c/b}$, and $visibility$ parameters (see Table \ref{tbl:parameter_ranges}). The CNN takes $64\times64$ pix images of three passbands as inputs and produces the aforementioned parameters as outputs. We simulated mock clusters as described in Sec. \ref{sec:mock_clusters}, placed them in randomly selected background fields of M83 and trained the network to distinguish between random stellar backgrounds and clusters, as well as to simultaneously infer cluster parameters. The pixel values of each passband image were individually normalized to the mean of zero and standard deviation of one to minimize the influence of photometric image calibrations. These were then rescaled with the arcsinh function.

The continuous $class_{c/b}$ parameter gives us an indication of a cluster's presence in an image. Unlike the other parameters, it does not have a strict physical interpretation, and it's meant as a way for the CNN to express its certainty on whether the processed image contains a cluster. In contrast, $visibility$ was developed as a proxy for signal-to-noise, and is defined as the ratio of a cluster's flux divided by the standard deviation of its background. This parameter lets us control the faintness of cluster candidates that we detect. Since $class_{c/b}$ works in the same way on all clusters regardless of their background or parameters, and the masses of our training clusters can go quite low, it becomes an unreliable measure when the clusters are very dim or drowned out by a dense background. Deciding whether objects like these should be included in a sample needs a separate source of information. For this purpose the $visibility$ parameter is used. Lowering the values at which we cut off both of these parameters allows us to detect more of genuine clusters at the cost of more false positives.

\subsection{Cluster search procedure}

For cluster search the network was applied on input images of fixed-size, which were obtained by sliding a $64\times64$ pix window through the whole mosaic with a step of 1 pixel. In every such window the CNN infered the parameters that indicate a cluster's presence (namely $class_{c/b}$ and $visibility$), as well as age, mass, extinction, and size. In effect, value-maps are obtained for the whole galaxy, where each pixel has an associated set of inferred parameters. Of particular note is the $class_{c/b}$ value-map, which can be interpreted as each pixel having an associated likelihood of there being a cluster centered on it. This approach is similar to \cite{2020ApJS..248...20H}, who perform source detection, segmentation and morphological pixel-by-pixel classification.

Star clusters of a wide variety of sizes exist and localizing their center precisely can be difficult. For sparser clusters, two nearby pixels can easily be of nearly the same $class_{c/b}$ value and both of these pixels can be valid interpretations of where the center is located. In such scenarios, while trying to localize a cluster, simply taking pixels with high $class_{c/b}$ values is not enough. Instead, we first smoothed the $class_{c/b}$ map with a Gaussian kernel ($\sigma=3$ pixels) and then searched for local peaks. This produces a large number of object candidates to work with.

Similarly, for each such object candidate, values of $visibility$, age, mass, extinction, and size parameters are obtained by taking several pixels around the peak of $class_{c/b}$ into account. This was done by computing a weighted average of the inferred parameters around a peak, using each pixel's $class_{c/b}$ value, as well as its distance from said peak in a $5\times5$ window, as the weight. This results in each object candidate having an associated set of inferred parameter values.

\subsection{Artificial cluster tests}

Artificial cluster tests were used to evaluate the CNNs detection performance. For that we generated $\sim$26,000 mock clusters and placed them on the mosaic in a regular grid. The result of this can be seen in Fig. \ref{fig:mocks_on_galaxy}. The mock clusters are spaced 64 pix apart from each other, covering the whole extent of the mosaic. Fig. \ref{fig:mocks_on_galaxy}b shows a sparser outskirt of the galaxy with an OB association. Most of the mock clusters can be easily seen in the sparser areas of this image, however, even bright clusters can be out-shined by the dense background of the OB association. Fig. \ref{fig:mocks_on_galaxy}c shows a central spiral-arm region of the galaxy, where fainter clusters are harder to see in the dense stellar background. The variety of mock cluster appearances are obtained due to the random sampling of cluster ages, masses, sizes, extinctions, as well the cluster's star masses and positions.

\begin{figure}
\centering
\includegraphics[width=1.0\columnwidth]{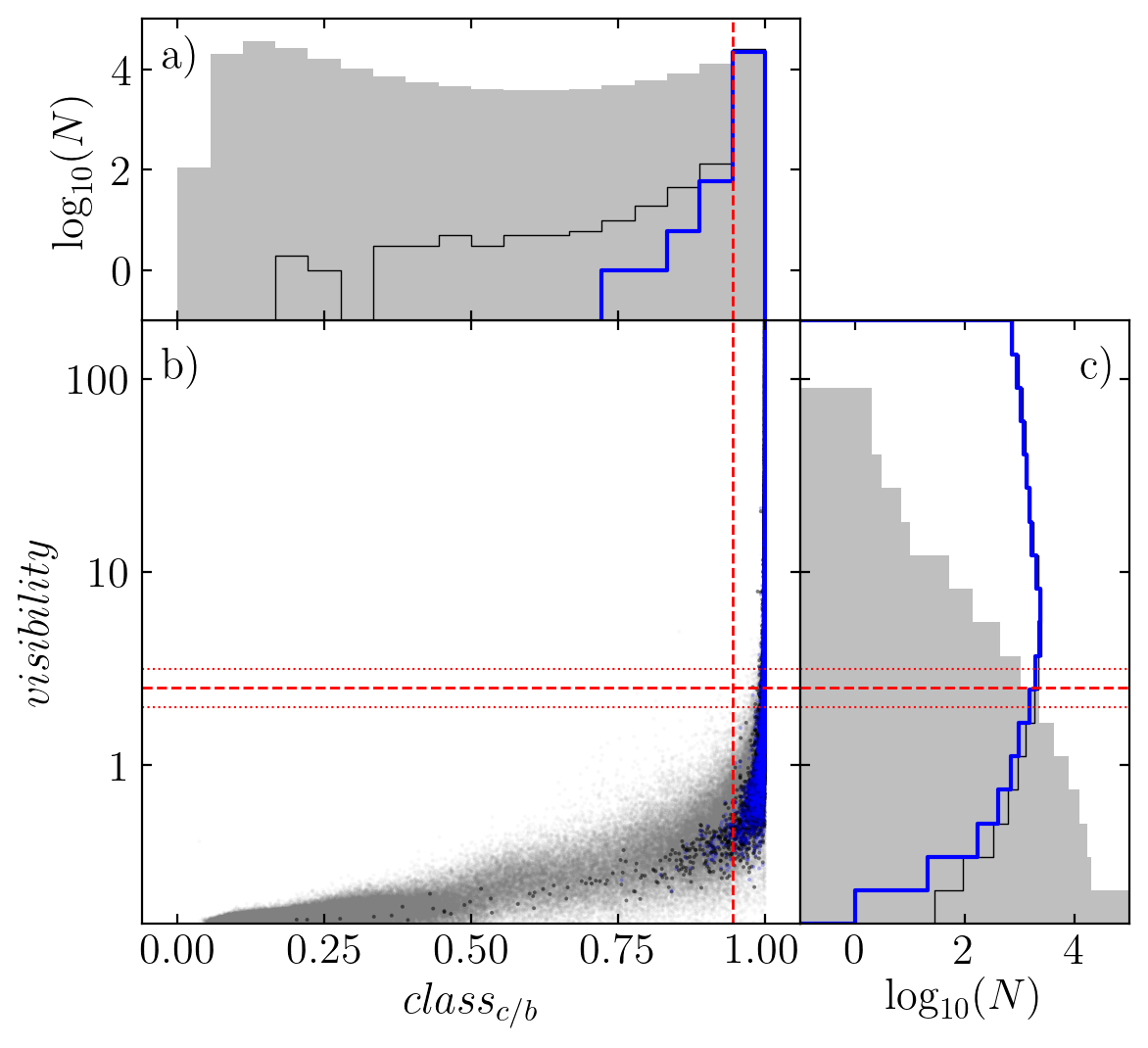}
\caption{Detection results of star clusters with a CNN: 1) on the mosaic of M83 (gray, $\sim$200,000 object candidates) and 2) on the artificial cluster test mosaic seen in Fig. \ref{fig:mocks_on_galaxy}a (blue, $\sim$23,000 total recalled objects, which matched with the input list with a maximum distance of 3 pixels). Recovered $\sim$24,600 mock clusters with a less strict maximum cross-matching distance of 6 pixels are displayed as black histograms and dots. The histograms are logarithmic counts of objects marginalized over $visibility$ (top) and $class_{c/b}$ (right). The dashed red line shows the chosen $visibility=2.5$ and $class_{c/b}=0.95$ thresholds, while the dotted red lines show $visibility=2$ and 3.}
\label{fig:matched_visibility_vs_clustericity}
\end{figure}

\begin{figure*}
    \centering
    \includegraphics[width=1.0\textwidth]{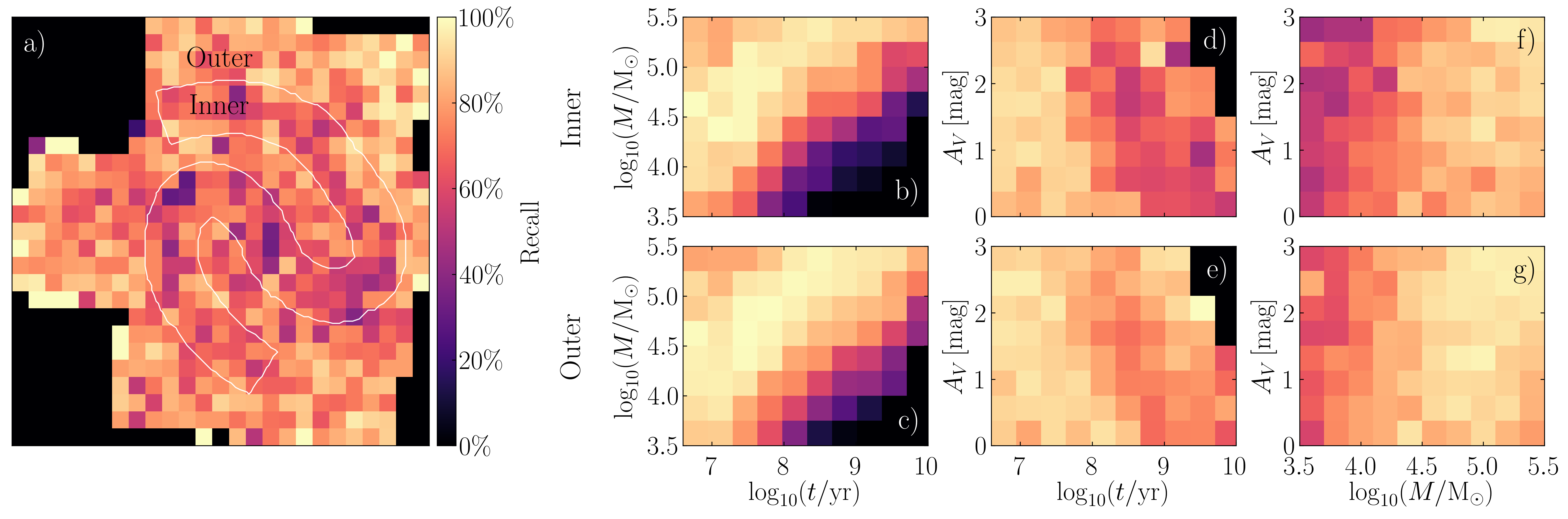}
    \caption{Recall (completeness) maps of artificial cluster tests. Each bin represents the number of objects that were recovered divided by the number of objects that were placed. Panel a shows recall of $\log_{10}(M/{\rm M_\odot})<4.5$ mock clusters as a function of their spatial position in the galaxy (as in Fig. \ref{fig:mocks_on_galaxy}), with the spiral arm regions outlined in white. Panels on the right show recall as a function of age and mass (b and c), age and extinction (d and e), as well as mass and extinction (f and g). The top row of panels (b, d, and f) show the recall within the spiral arms (inside the white outline in panel a), while the bottom row of panels (c, e, and g) show recall outside of the spiral arms. Note, that the black bins signify areas where no artificial clusters were inserted.}
    \label{fig:recalled_cluster_params}
\end{figure*}

We performed the search for artificial clusters on the mock-cluster-filled mosaic and cross-matched the potential objects with the inserted mock cluster catalog, using a maximum distance of 3 pixels, in order to capture the mocks with a high confidence. This resulted in $\sim$23,000 objects displayed as blue points in Fig. \ref{fig:matched_visibility_vs_clustericity}. For reference, another sample of objects are displayed as black points and histograms, using a maximum distance of 6 pixels when crossmatching, which results in $\sim$24,600 objects. We repeated the same search procedure on the original M83 mosaic and found $\sim$200,000 object candidates with various $class_{c/b}$ and $visibility$ values visualized as gray points in Fig. \ref{fig:matched_visibility_vs_clustericity}, which contain real cluster candidates we aim to select later.

In Fig. \ref{fig:matched_visibility_vs_clustericity} it can be seen that the majority of recalled mock clusters have high $class_{c/b}$ and $visibility$ values, but that is also true for a number of the original-mosaic detections. This is expected, as the gray points include real clusters of the M83 galaxy. We vary the thresholds for $class_{c/b}$ and $visibility$ to analyse recall as a function of spatial position in the galaxy and the mock cluster parameters in Fig. \ref{fig:recalled_cluster_params}.

Fig. \ref{fig:recalled_cluster_params} shows recall maps (fraction of clusters that were found out of those that were inserted) of artificial clusters after applying the $class_{c/b}$ threshold of 0.95 and the $visibility$ threshold of 2.5. In Fig. \ref{fig:recalled_cluster_params}a the recall is shown spatially, for clusters with $\log_{10}(M/{\rm M_\odot})<4.5$. The spiral arm areas used for recall analysis are outlined in white. This divides the galaxy into two regions: inner (i.e. the area on the arms) and outer. For the inner regions the recall of the CNN is lower than for the outer regions, which is expected due to the relatively higher stellar background in and around the arms. A more detailed breakdown of recall is given in panels Fig. \ref{fig:recalled_cluster_params}b-g.

In Fig. \ref{fig:recalled_cluster_params}b and c the recall of clusters as functions of age and mass is shown. For both the inner and outer regions of the spiral arms recall decreases with older ages and lower masses. The effect is more pronounced for the inner region, because in a dense background faint clusters become much harder to find. Fig. \ref{fig:recalled_cluster_params}d and e show recall as a function of age and extinction. The recall is broadly uniform, however, for the inner region lower values of recall can be observed for old and high extinction clusters. Fig. \ref{fig:recalled_cluster_params}f and g show recall as a function of mass and extinction. The decrease of recall occurs for low luminosity clusters of low masses and high extinctions. This is more pronounced on the spiral arms.

\section{Results} \label{sec:results}

\subsection{Cluster selection}

\begin{figure*}
\centering
\includegraphics[width=0.9\textwidth]{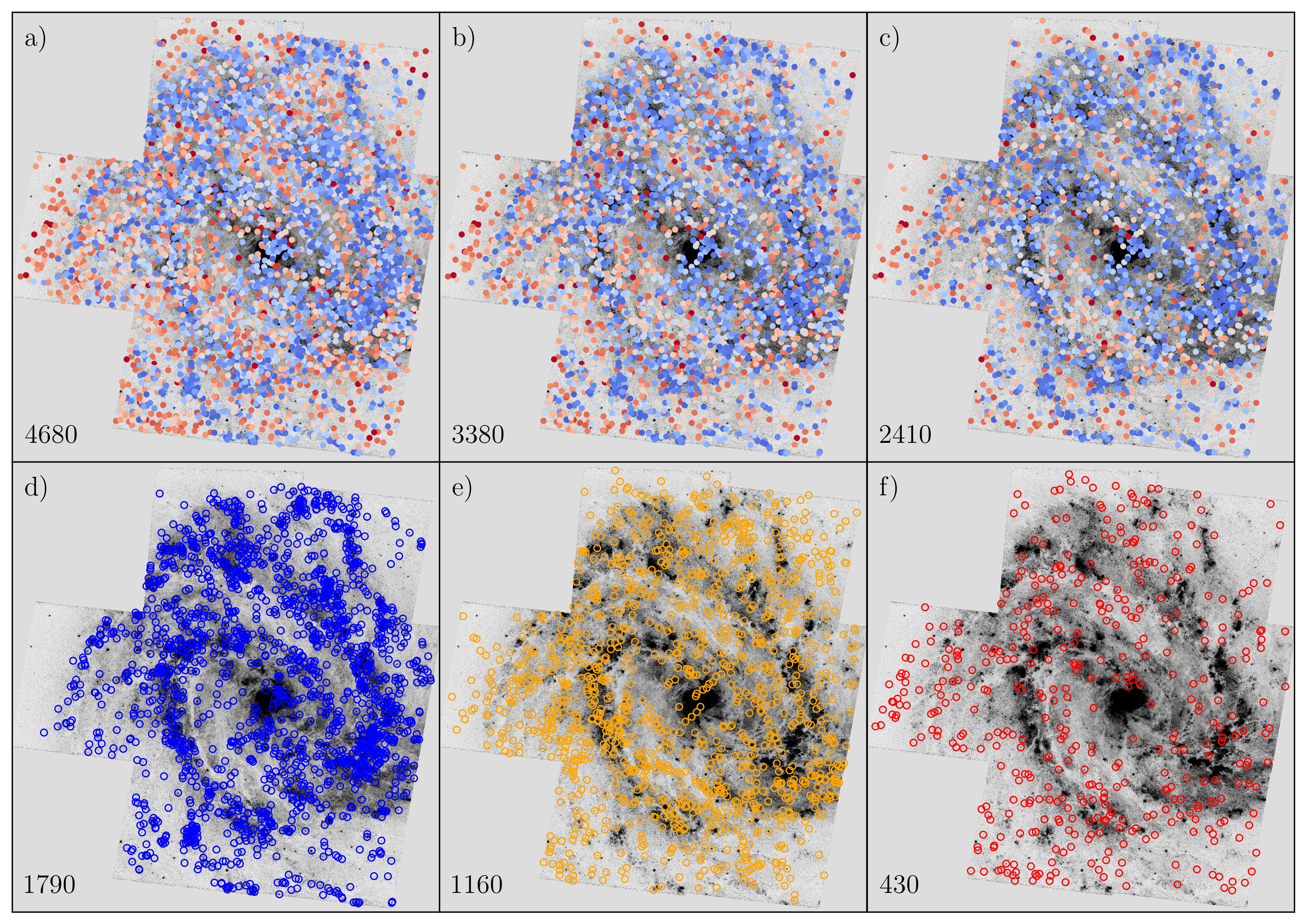}
\caption{Detected cluster candidates on the M83 mosaic. The top row of panels depict different $visibility$ thresholds used to filter out candidates: a) 2, b) 2.5, and c) 3, with blue indicating young clusters $\log_{10}(t/{\rm yr})=6.6$ and red -- old clusters $\log_{10}(t/{\rm yr})\geq9$. The bottom row of panels depict clusters from panel b, split into ages of: d) $\log_{10}(t/{\rm yr})<7.5$, e) $7.5\leq\log_{10}(t/{\rm yr})<8.5$, and f) $8.5\leq\log_{10}(t/{\rm yr})<10.1$. The number of cluster candidates is displayed on the lower left of each panel.}
\label{fig:age_cuts_on_2d_galaxy}
\end{figure*}

The $\sim$200,000 object candidates detected on the original M83 mosaic, depicted as gray dots in Fig. \ref{fig:matched_visibility_vs_clustericity}, were filtered using a $visibility$ threshold of 2 and $class{c/b}$ of 0.95. This resulted in 5,460 object candidates, which were inspected visually and obvious false positives were removed. This process left 4,680 cluster candidates with $visibility>2$ and 3,380 cluster candidates with our baseline threshold of $visibility>2.5$. The visual inspection of possibly missed cluster candidates with $visibility>2.5$ was estimated to be in line with artificial cluster tests seen in Fig. \ref{fig:recalled_cluster_params}, however we did not include any human-detected clusters and analyse only CNN-detected candidates for consistency.

Fig. \ref{fig:age_cuts_on_2d_galaxy} shows these candidates on the M83 mosaic using several $visibility$ thresholds, as well as highlighting the spatial distributions of cluster ages. The top row of panels in Fig. \ref{fig:age_cuts_on_2d_galaxy} depict different $visibility$ thresholds, with panel b representing our baseline threshold, denoted as a red dashed line in Fig. \ref{fig:matched_visibility_vs_clustericity}. Increasing or decreasing the threshold by 0.5 produces a difference of $\sim$30\% in the number of candidate counts.

The bottom row of panels in Fig. \ref{fig:age_cuts_on_2d_galaxy} depict clusters from panel b, grouped by age. Young cluster candidates seen in panel d are concentrated around the spiral arms and other star forming regions, while intermediate and older clusters in panels e and f are spread around relatively uniformly, with a somewhat higher concentration around spiral arms for the intermediate age clusters.

\subsection{Inferred parameters}

\begin{figure*}
\centering
\includegraphics[width=0.9\textwidth]{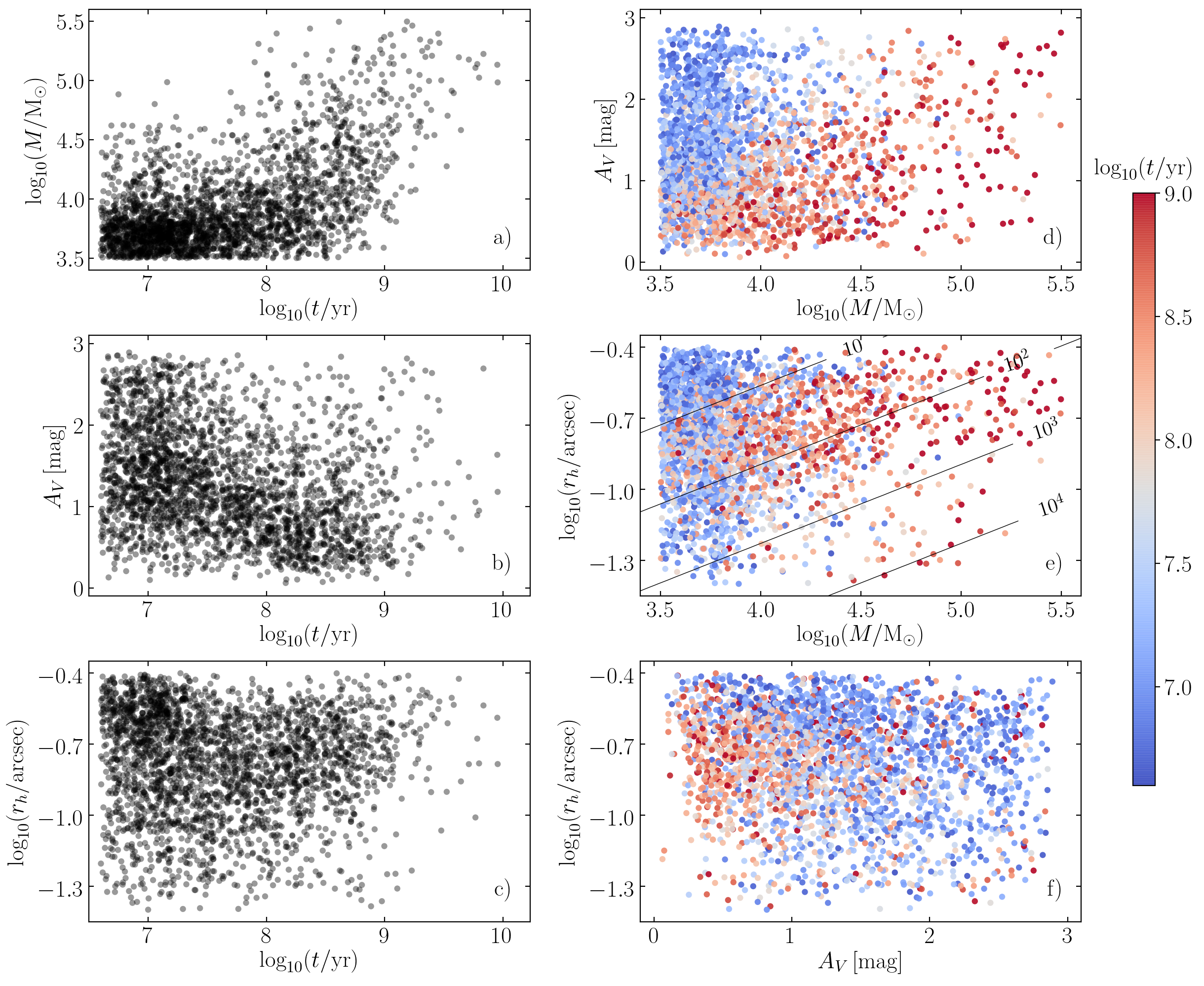}
\caption{Parameter distributions of 3,380 cluster candidates detected by CNN on the M83 mosaic. Panels show combinations of: a) mass vs. age, b) extinction vs. age, c) size vs. age, d) extinction vs. mass, e) size vs. mass, and f) size vs. extinction. The colors in panels d, e, and f represent cluster ages, with blue indicating young clusters $\log_{10}(t/{\rm yr})=6.6$ and red -- old clusters $\log_{10}(t/{\rm yr})\geq9$. The diagonal lines in panel e represent various cluster density $\rho_h/{\rm (M_\odot \cdot pc^{-3})}$ levels.}
\label{fig:param_param_detections}
\end{figure*}

Fig. \ref{fig:param_param_detections} shows the inferred age, extinction, mass, and size distributions of the cluster candidates. Diagonal cutoffs due to the $visibility$-limited detection can be seen in Fig. \ref{fig:param_param_detections}a for old low-mass cluster and in Fig. \ref{fig:param_param_detections}b for old high-extinction clusters. This was also seen for artificial clusters in Fig. \ref{fig:recalled_cluster_params}. In addition, because of age-extinction degeneracies, noted in Paper II, arising due to the passbands that were used, parameter inference for old clusters can be unreliable.

\begin{figure*}
\centering
\includegraphics[width=0.8\textwidth]{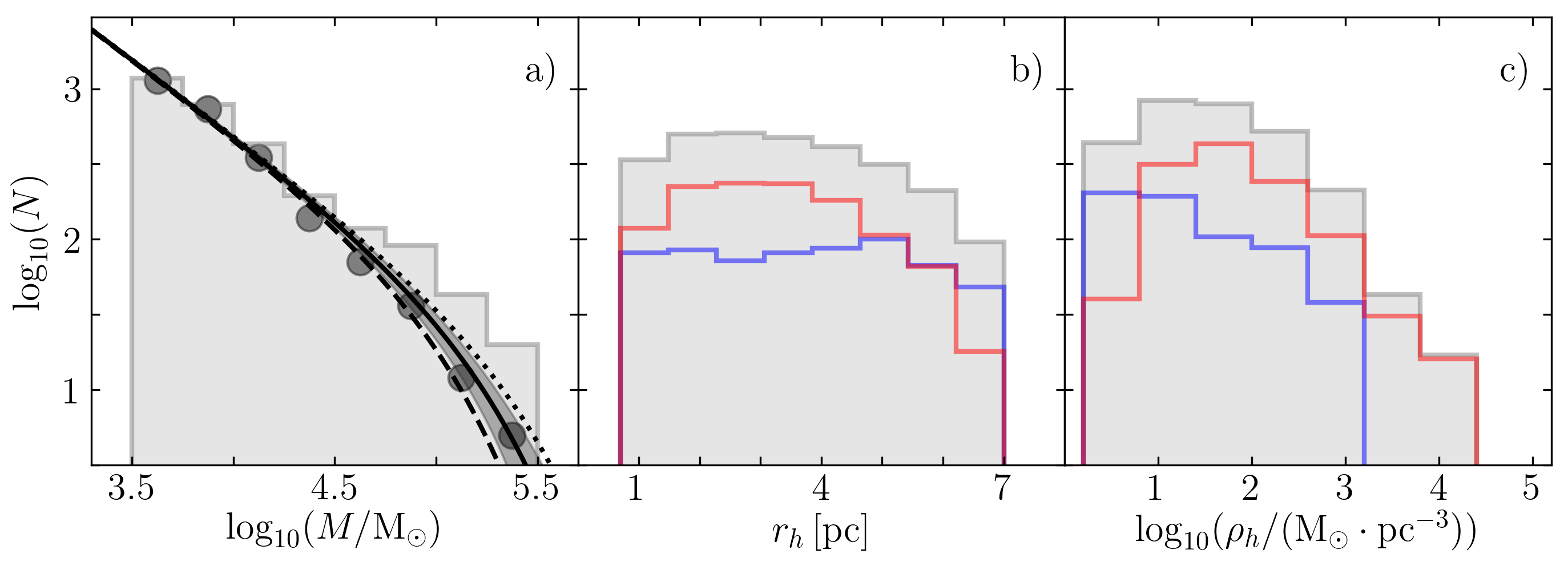}
\caption{Mass (a), size (b), and density (c) histograms of the 3,380 cluster candidates. In panel a circles represent the distribution with age $\log_{10}(t/{\rm yr})<8.5$, while the shaded gray histogram represents the whole cluster sample. Lines represent the Schechter mass distribution functions of the form $dN/dM=A \cdot M^{-2} \cdot \exp(-M/M_*)$, where $M_*$ is: 1) $10^5$ (dotted line), 2) $1.6 \cdot 10^5$ $M_\odot$ (solid line, with the shaded area encompassing its Poisson standard deviation), and 3) $2.5 \cdot 10^5$ $M_\odot$ (dashed line). In panels b and c the gray shaded histograms show the whole cluster sample, the blue outlines -- young $\log_{10}(t/{\rm yr})<7$ and red -- older $\log_{10}(t/{\rm yr})\geq7.7$ clusters.}
\label{fig:mass_histograms}
\end{figure*}

In Paper II we have compared the CNN inference results to catalogs of \cite{2011MNRAS.417L...6B}, \cite{2015MNRAS.452..525R}, and \cite{Harris_2001} showing a reasonably good agreement between the catalogs and the CNN results. The parameter distributions of cluster candidates seen in Fig. \ref{fig:param_param_detections} also corresponds well to the \cite{2011MNRAS.417L...6B} sample, inferred with the CNN and displayed in Fig. 18 of Paper II.

However, a numerous population of young-age low-mass extended cluster candidates (which by visual inspection could be classified as small associations of stars), is clearly visible in the bottom-left of Fig. \ref{fig:param_param_detections}a and the top-left of Fig. \ref{fig:param_param_detections}c. These are low density objects (Fig. \ref{fig:param_param_detections}e), with many of them having density values of $\rho_h<100\,{\rm M_\odot \cdot pc^{-3}}$. Mass and size relations of clusters in various galaxies are summarized in Fig. 9 of \cite{2019ARA&A..57..227K}, indicating that the majority of objects follow a $\rho_h\sim100\,{\rm M_\odot \cdot pc^{-3}}$ line and that the ranges of density are consistent with CNN results. 

Although the definition of what observationally should be considered a cluster vs. an unbound association is not clear at such young ages \citep{2012MNRAS.419.2606B}, these objects are detected by our CNN as cluster candidates. Comparing the cluster appearance of different catalogs one can see a variety of inclusion criteria and in our case this is decided by the CNN and the parameter ranges of mock cluster on which it was trained.

In addition, in both \cite{2012MNRAS.419.2606B} Fig. 14 and \cite{2015MNRAS.452..525R} Fig. 3 a trend can be seen connecting cluster age and size. We can also see a similar effect in Paper II Fig. 18, where parameters of the same catalog are derived via CNN. However, this is not present in our cluster candidate sample used for this paper, as seen in Fig. \ref{fig:param_param_detections} panel c, where a large number of young low-mass extended objects are observed.

\cite{2010ARA&A..48..431P} discussed the importance of detection limits and their influence on the lack of low mass clusters in catalog construction. This is explored in Fig. \ref{fig:mass_histograms}a, where circles represent the mass distributions with ages $\log_{10}(t/{\rm yr})<8.5$. The superimposed lines are Schechter type mass functions \citep{2010ARA&A..48..431P}, of the form $dN/dM=A \cdot M^{-2} \cdot \exp(-M/M_*)$, with different amounts of truncation, which were previously derived by \cite{2012MNRAS.419.2606B} on their sample of M83 clusters. The Schechter type function matches the data, explaining the numerous low-mass clusters seen in Fig. \ref{fig:param_param_detections} and are consistent with \cite{2020ApJ...893..135M}.

In Fig. \ref{fig:mass_histograms}b we show the distributions of sizes of young and older cluster candidates. The sizes are somewhat evenly distributed, with a drop for smaller and larger clusters. This distribution is similar to the one found for Galactic clusters based on Gaia observations by \cite{2020MNRAS.495.2882S} and consistent with previous results for M83 clusters \citep{2015MNRAS.452..525R}. In our sample, older clusters are smaller than younger clusters, which is as expected due to the fact that extended clusters are disrupted more easily due to various factors such as galactic tidal forces and passing through molecular clouds in the spiral arms. These effects can also be seen in Fig. \ref{fig:mass_histograms}c, with older clusters being more dense than younger clusters.

\subsection{Distance from spiral arms} \label{sec:distance_from_spiral_arms}

According to the density wave theory \citep{2016ARA&A..54..667S}, the spiral arms of galaxies are structures that occur as objects and gas move into them, slow down, and then move out again. In this model the spiral pattern itself rotates at a fixed angular velocity, while objects, such as star clusters, move at differing angular velocities, which depend on their distance to the galactic center. In this scenario one would expect star clusters inside the corotation radius to outpace the spiral arms, creating an age gradient with older clusters being observed further away from the arms than younger clusters. This is supported by observations, such as \cite{2019ApJ...874..177M}, who measured the pitch angle of spiral arms for a sample of galaxies at several wavelengths, finding, that it is systematically larger in blue and smaller in red passbands, claiming the results to be consistent with enhanced stellar light located downstream of a star forming region for a galactocentric radius smaller than the corotation radius. \cite{2018PASJ...70..106S} have identified bow shock structures as wavy arrays of star forming regions distributed along the spiral arms of M83 and used the shape and orientation of HII region shock cones to estimate the position of the corotation radius coinciding to the one estimated by \cite{2014PASJ...66...46H}.

To analyze the distribution of parameters as a function of the distance from the spiral arms, we traced out dust lanes along the inner part of the arms as a reference (shown as black curves in Fig. \ref{fig:parameters_on_2d_galaxy}a). This follows the approach of \cite{2018MNRAS.478.3590S}, where cluster distances from dust lanes were used to analyze age gradients in three spiral galaxies. However, we used linear instead of azimuthal distances due to projection effects and the possibly warped disk of M83 (see Section \ref{sec:discussion} for more details).

\begin{figure}
\centering
\includegraphics[width=1.0\columnwidth]{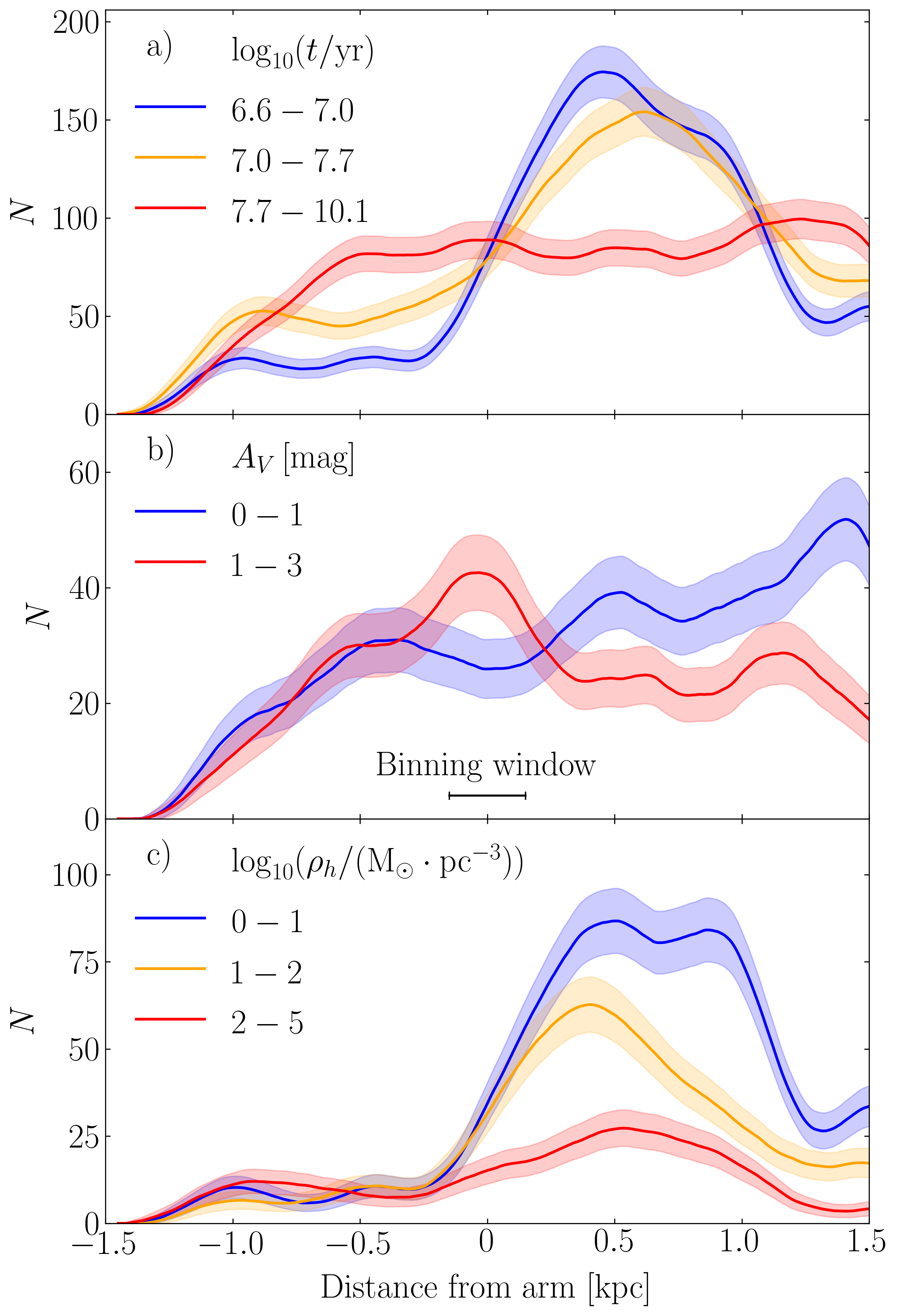}
\caption{Counts of the selected cluster candidates as a function of their distance from the spiral arms of the galaxy with: a) 3 different cuts for age, b) 2 cuts for extinction for clusters with $\log_{10}(t/{\rm yr})>8$, and c) 3 cuts for density for clusters with $\log_{10}(t/{\rm yr})<7$. The binning window line on the bottom of panel b represents the bin width used to calculate the cluster count at each distance point. Negative distances indicate clusters trailing the spiral arms, while positive distances indicated clusters leaving and outpacing the arms. Poisson error bars are visualized as shaded areas around each curve.}
\label{fig:histograms_as_distance_to_arm}
\end{figure}

Fig. \ref{fig:histograms_as_distance_to_arm} shows the counts of cluster candidates as a function of their shortest-line distances from the dust lanes of the spiral arms of the galaxy (see Fig. \ref{fig:parameters_on_2d_galaxy}a). Note, that for trailing (negative distance) cluster candidates we only collect objects up to about $\sim$1 kpc, before running into the leading (positive distance) cluster candidates of a different spiral arm (see Fig. \ref{fig:age_cuts_on_2d_galaxy}a) and vice versa. This has to be kept in mind as any trends beyond those points will exhibit boundary effects, limiting the interpretation of this data within $\pm$1 kpc.

Fig. \ref{fig:histograms_as_distance_to_arm}a shows 3 different cuts of ages, with blue indicating $\log_{10}(t/{\rm yr})<7$, orange -- $7\leq\log_{10}(t/{\rm yr})<7.7$, and red -- $\log_{10}(t/{\rm yr}) \geq 7.7$. The blue and orange lines represent recently formed clusters, associated with the spiral arm, while the red line shows the field population. The peak for young clusters appears closer to the spiral arms, at $\sim$0.4 kpc in the leading direction, while for the oldest field population of clusters it stays approximately flat. This corresponds well to the result obtained for other spiral galaxies in \cite{2018MNRAS.478.3590S}.

Assuming the rotation curve speed of $\sim$160 km/s \citep{2016MNRAS.462.1238H} and the spiral pattern speed of $\sim$110 km/s at the distance of $\sim$2.5 kpc \citep{2004ApJ...607..285Z}, where the majority of our cluster candidates (that are in the vicinity of the spiral arms) are located, we expect objects to overtake and lead the arms at the rate of $\sim$50 pc per one million years. This is in good agreement with the peak we find of $6.6\leq\log_{10}(t/{\rm yr})<7$ cluster candidates at $\sim$0.4 kpc and $7\leq\log_{10}(t/{\rm yr})<7.7$ cluster candidates at $\sim$0.7 kpc. Note, also, that cluster candidates of $\log_{10}(t/{\rm yr}) \approx 8$ have already completed a full orbit around the galaxy, which supports our lower age cut of $\log_{10}(t/{\rm yr})\geq7.7$ as representing the field population of clusters.

Fig. \ref{fig:histograms_as_distance_to_arm}b shows the extinctions of older clusters, with ages of $\log_{10}(t/{\rm yr})>8$. This age cut was chosen to select a sample of clusters that are already independent of the regions where they were formed, giving us an insight into the extinctions of the field population. The blue line indicates low extinction objects with $A_V<1$ mag, while the red line indicates high extinction objects with $A_V\geq1$ mag. The peak of the number of objects with high extinction occurs right on top of the spiral arm dust lanes. While the number of low extinction objects gradually increases leading the arms, reaching the highest counts at the distance of $\gtrsim$1 kpc in front of the arms. This corresponds to objects which are located in sparse areas of the arms and are, therefore, less obscured by dust.

Fig. \ref{fig:histograms_as_distance_to_arm}c shows the densities of the population of younger clusters, with ages of $\log_{10}(t/{\rm yr})<7$ (corresponding to the blue line of panel a). This younger age cut was chosen to get the densities of clusters that were formed within the nearby spiral arm instead of being in the field population. The blue line indicates densities of $\log_{10}(\rho_h/{\rm (M_\odot \cdot pc^{-3})})<1$, the orange line -- $1\leq\log_{10}(\rho_h/{\rm (M_\odot \cdot pc^{-3})})<2$, and the red line -- $\log_{10}(\rho_h/{\rm (M_\odot \cdot pc^{-3})})\geq2$. There is a vague trend of higher density cluster counts peaking further away from the spiral arms than the lower density objects.

\begin{figure}
    \centering
    \includegraphics[width=1.0\columnwidth]{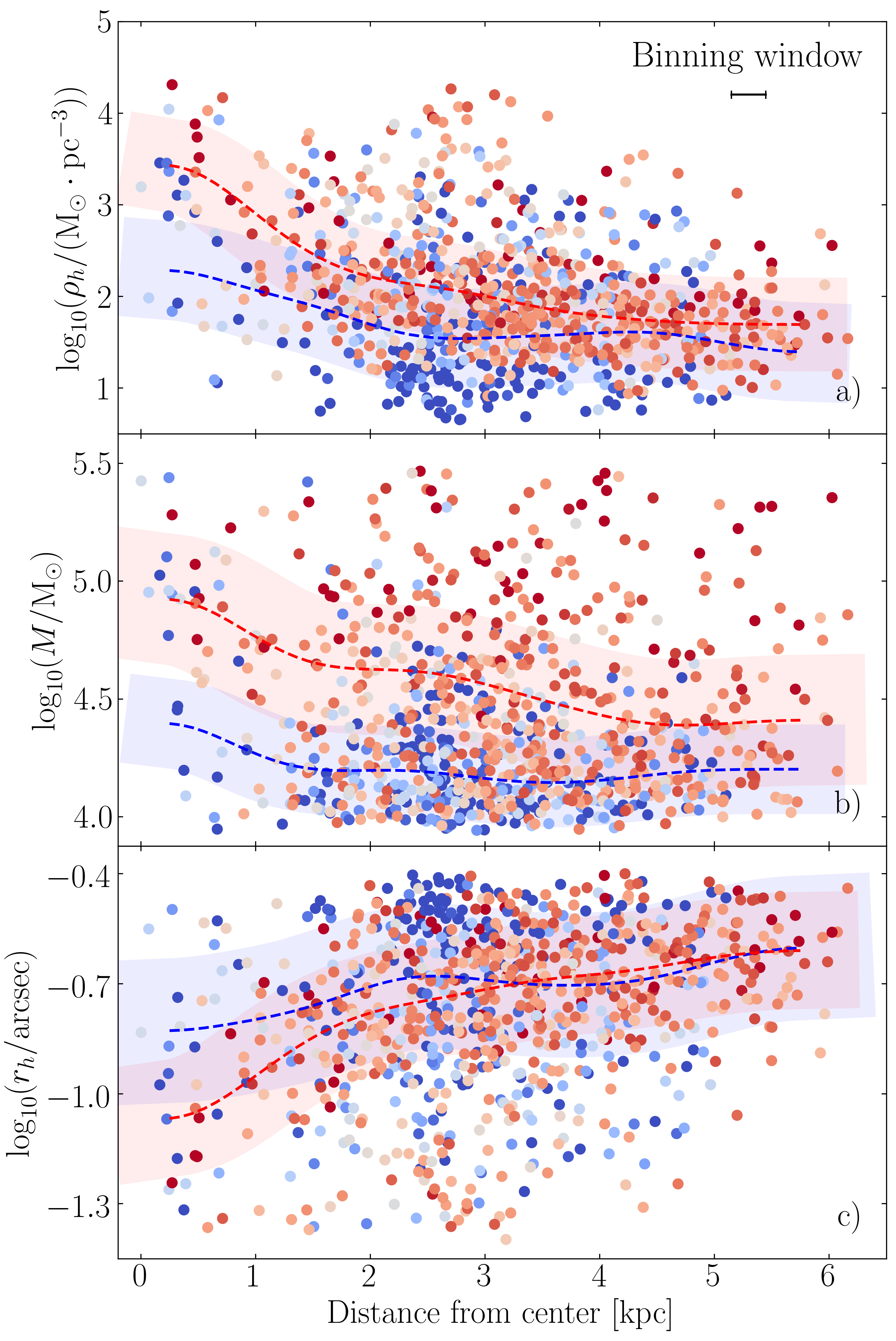}
    \caption{Densities (a), masses (b), and sizes (c) of the selected cluster candidates vs. their distance from the galactic center. Colors of the circles indicate ages, starting with $\log_{10}(t/{\rm yr})\leq7$ (blue) and up to $\log_{10}(t/{\rm yr})\geq9$ (red). The two dashed lines represent the mean parameter values for young ($\log_{10}(t/{\rm yr})<7.5$) and old ($\log_{10}(t/{\rm yr})\geq8.5$) cluster candidates, with the shaded area around the dashed lines representing the standard deviation of each parameter. The binning window line in the top right of panel a represents the width of the window used to calculate the means at each distance point.}
    \label{fig:distance_vs_concentration}
\end{figure}

\begin{figure*}
    \centering
    \includegraphics[width=0.85\textwidth]{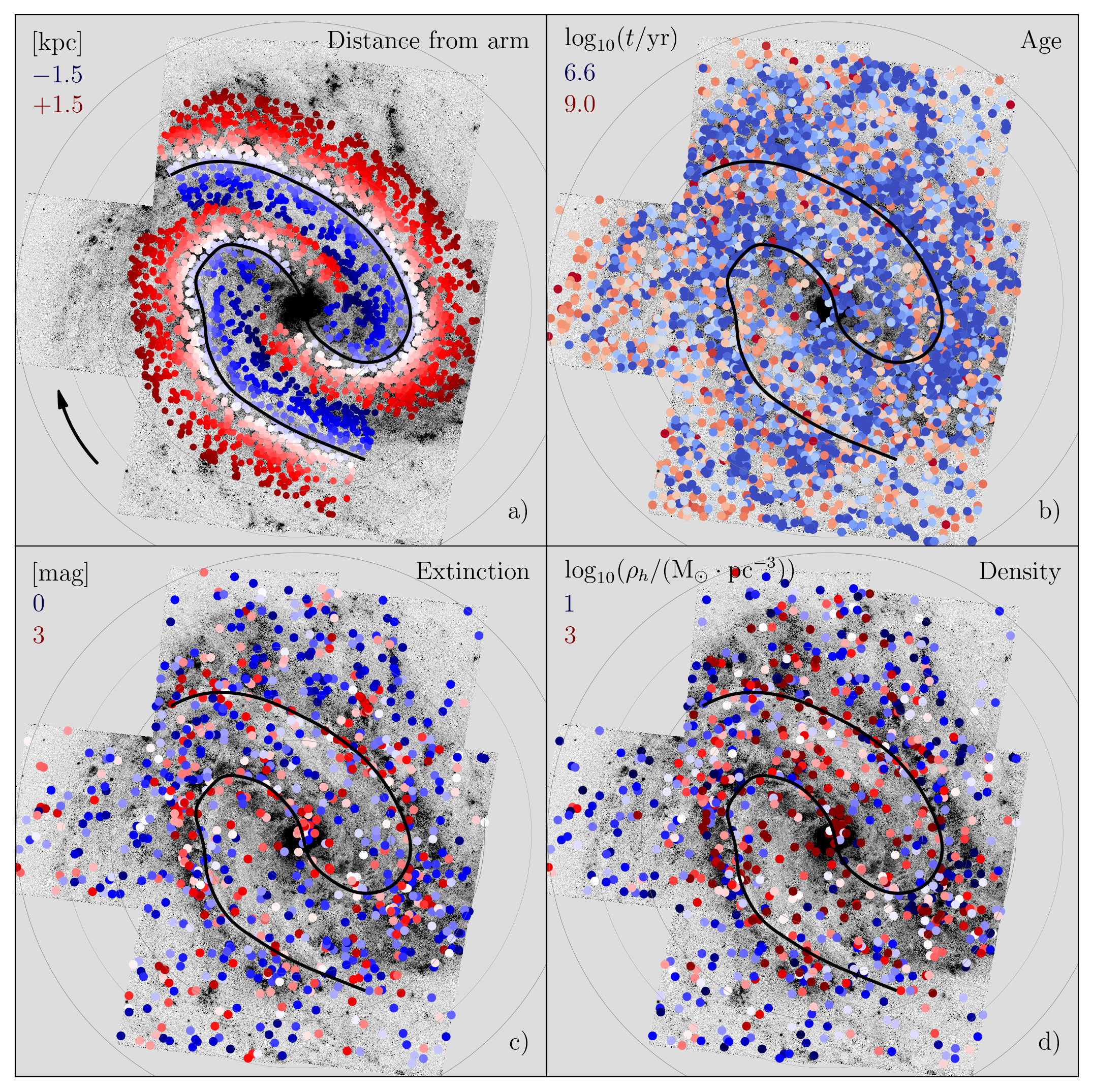}
    \caption{Spatial distribution of selected cluster candidates displayed on the F438W band. Panel a) shows the distance of cluster candidates from their nearest spiral arm as an increasing intensity of red for the leading arm areas, and blue for the trailing arm areas. Clusters are drawn up to the distance of 1.5 kpc. Panel b) shows the ages of clusters starting with $\log_{10}(t/{\rm yr})=6.6$ (blue) and up to $\log_{10}(t/{\rm yr})\geq9$ (red). Panel c) shows extinctions of $\log_{10}(M/{\rm M_\odot})>4$ cluster candidates, with the color map showing clusters starting at $A_V=0$ mag (blue) up to $A_V=3$ mag (red). Panel d) shows the same sample of clusters as panel c), only with the color coding corresponding to density values starting with $\log_{10}(\rho_h/{\rm (M_\odot \cdot pc^{-3})})<1$ (blue) and going up to $\log_{10}(\rho_h/{\rm (M_\odot \cdot pc^{-3})})\geq3$ (red). The faint concentric circles indicate distances from the galactic center, starting with 1 kpc and up to 6 kpc, with a step of 1 kpc. The arrow in the bottom left of panel a indicates the direction of rotation; the galactic disc is inclined with the NW part being closer to observer.}
    \label{fig:parameters_on_2d_galaxy}
\end{figure*}

To verify that these trends did not arise due to selection effects, we took the mock clusters that were inserted in a uniform grid on the whole galaxy image, recovered them with the CNN and sliced their inferred parameter ranges in the same way as was done for Fig. \ref{fig:histograms_as_distance_to_arm}. This way both the counts and the relevant parameters of the clusters are distributed uniformly over the galaxy and should result in curves without the peaks seen for real clusters. For these recovered mock clusters the ends of the distributions (for low negative and high positive distances) do tend to have a lower number of clusters, due to boundary effects of the maximum possible distance from the spiral arms. For the area in which we draw our results (approximately between $-0.2$ and 1.4 kpc) the recovered mock cluster distributions are flat. For age in particular, the recovered mock young cluster distributions are very similar to the real old cluster distributions, which is exactly what we would expect due to the old real cluster population being relatively uniformly distributed in the galaxy. Selection effects should not have a measurable impact on the trends found for cluster candidates.

\subsection{Distance from galaxy center}

In Fig. \ref{fig:distance_vs_concentration}a it can be seen that older clusters are more dense closer to the galactic center, while for young clusters this effect is less pronounced. As density is a function of mass and size, we show their distributions in Fig. \ref{fig:distance_vs_concentration}b and c. Both young and old clusters are more massive near the galactic center, possibly due to the fact that lower mass clusters break apart more easily in a denser environment. However, the average mass differences between these two populations get smaller the further away they are from the center. It can also be seen in Fig. \ref{fig:distance_vs_concentration}c, as there are more compact clusters near the center of the galaxy, with the effect being more pronounced for the older population. This is in agreement with \cite{2017MNRAS.468.1769F}, who have found that the surface density of molecular clouds varies with distance from the galactic center and shown that the mass of the most massive cluster decreases with distance.

We have also repeated the mock cluster test described in Section \ref{sec:distance_from_spiral_arms} to check whether selection effects have an impact on the results presented in this section as well and found no systematic biases.

\subsection{Spatial distribution}

Fig. \ref{fig:parameters_on_2d_galaxy} shows the spatial distribution of selected cluster candidates. The dust-band-traced spiral arms used for the analysis in Fig. \ref{fig:histograms_as_distance_to_arm} are displayed as black curves. In Fig. \ref{fig:parameters_on_2d_galaxy}a a fraction of the selected cluster candidates are displayed with colors corresponding to their distance from the spiral arms, up to $\pm$1.5 kpc. The color blue indicates clusters trailing the arm, seen as negative distances in Fig. \ref{fig:histograms_as_distance_to_arm}, and the color red indicates clusters leading the arm.

Fig. \ref{fig:parameters_on_2d_galaxy}b shows the ages of the cluster candidates. A higher concentration of young clusters on the spiral arms and other star forming regions is observed, while older clusters are more evenly spread out in-between the arms and in the outer regions of the galaxy. \cite{2012ApJ...753...26K} analyzed the distributions of ages of resolved stars and found groupings of young stars along a spiral arm and more evenly distributed old stars. This closely corresponds to the distributions of our cluster candidate sample in the same region.

Fig. \ref{fig:parameters_on_2d_galaxy}c shows the extinction of clusters more massive than $\log_{10}(M/{\rm M_\odot})>4$. There are more clusters with high extinction located on the spiral arms and other dust-rich areas of the galaxy. This can also be seen as a function of distance to the arm in Fig. \ref{fig:histograms_as_distance_to_arm}b.

Fig. \ref{fig:parameters_on_2d_galaxy}d shows the densities of the same higher-mass cluster sample as in Fig. \ref{fig:parameters_on_2d_galaxy}c. Higher density clusters concentrate around the spiral arms of the galaxy and closer to its central region in general, as is shown in Figs. \ref{fig:histograms_as_distance_to_arm}c and Fig. \ref{fig:distance_vs_concentration}a and could be due to less dense clusters being broken up more easily in these crowded environments.

CNN results allowed us to identify populations of clusters with different properties on and within the spiral arms, as well as get consistent results of the age gradient of clusters within the corotation radius of M83. Such data could be used to test and constrain various hypothesis of cluster formation and disruption by comparing it to models of galaxy evolution.

\section{Discussion} \label{sec:discussion}

Here we discuss the limitations of the method, as well as some of the caveats of our results and other relevant observations. The items are presented in a random order.

\begin{itemize}
    \item {\bf Detection threshold.} The $class_{c/b}$ and $visibility$ parameter thresholds can be varied freely to select a different number of cluster candidates, as can be seen in Fig. \ref{fig:matched_visibility_vs_clustericity}. We repeated our experiments using cluster candidate samples obtained with different thresholds and observed no notable differences in the results. The distributions seen in Figs. \ref{fig:mass_histograms}--\ref{fig:parameters_on_2d_galaxy} remain consistent with the presented analysis.

    \item {\bf Parameter ranges.} They set the limits within which the CNN can produce predictions. Of particular note is cluster mass, which was constrained to $3.5\leq\log(M/{\rm M_\odot})\leq5.5$, to cover the majority of M83 disc cluster population and is consistent with the mass ranges used by \cite{2018MNRAS.478.3590S}.

    \item {\bf Mosaic area.} Our sample is effectively limited to the area within approximately 6 kpc from the center of the galaxy by the 7 WFC3 fields. However, this spatial extent is sufficient for our analysis and is consistent with the three galaxy coverage used by \cite{2018MNRAS.478.3590S} for a similar study.

    \item {\bf Possible interaction.} A deep photographic image of M83 \citep{1997PASA...14...52M} has revealed ``an enormous loop around NW quadrant of the galaxy''. Also, radio observations of its outskirts show a significant extent of a disturbed nature, suggesting a possible retrograde interaction \cite{2016MNRAS.462.1238H}. This is possibly due to a closest approach 1--2 Gyr ago of a neighbouring galaxy \citep{2014ccds.book.....F}. In addition, by tracing the dust lanes as displayed in Fig. \ref{fig:parameters_on_2d_galaxy} we observed that the north-western spiral arm is slightly closer to the center than the south-eastern arm. This, in addition with the galaxy's large estimated inclination angle of $\sim$40 deg and its uncertainty \citep{2016MNRAS.462.1238H}, compelled us to measure cluster distances perpendicularly to the dust lanes of the arms and not along the direction of rotation as was done in \cite{2018MNRAS.478.3590S}.

    \item {\bf Detection reliability.} \cite{2017MNRAS.472.1315W} have demonstrated that in combination human and CNN-based machine classification of transient objects outperforms either one individually; humans tend to have lower false positive rates, but higher missed detection rates, while machine classification has the opposite trend. Therefore, we visually inspected the CNN-detected object candidates, removed possible false positives from the sample and counted objects potentially missed by the CNN. We estimated the false negative rate to be consistent with our artificial cluster tests and those of previous studies of limited samples by \cite{2011MNRAS.417L...6B} and \cite{2015MNRAS.452..525R}. Visual inspection also revealed that likely background galaxies (red extended objects) are not detected by the CNN as clusters. Note, \cite{2014MNRAS.440L.116S} compiled $\sim$1,800 clusters and associations of $\log_{10}(t/{\rm yr}) \lesssim 8.5$ and $\log(M/{\rm M_\odot}) \gtrsim 3.7$. We find $\sim$1,600 cluster candidates within these age and mass ranges.

    \item {\bf Sample selection.} From our experience in doing visual selection of star clusters in the Andromeda galaxy, based on Subaru telescope \cite{2008ApJS..177..174N} observations, which effectively give a similar resolution as HST does in M83, it was noted, that higher selection thresholds are applied by humans to avoid including too many faint clusters. This results in a limited sample of low-mass clusters deviating from Schechter mass functions as shown in \cite{2009ApJ...703.1872V}. A similar effect is seen in the catalog of \cite{2011MNRAS.417L...6B} as shown in of Paper II. Examples of mock clusters in Paper II reveal that such low visibility objects might be easily rejected from a visually compiled sample, but could be genuine star clusters.

    \item {\bf Young low-mass candidates.} Comparing Fig. \ref{fig:param_param_detections}a with Fig. 14a of Paper II, the high number of young low-mass cluster candidates of this work does not significantly overlap with random background inference results, which are concentrated at $\log_{10}(t/{\rm yr})>7.5$ and $\log(M/{\rm M_\odot}) < 3.75$. Even though in Fig. 14e of Paper II a high number of backgrounds are identified as low-mass extended objects, they are not concentrated around young ages $\log_{10}(t/{\rm yr})<7$, while our cluster sample exhibiting the same properties in Fig. \ref{fig:param_param_detections}e are younger objects. In addition, in Fig. 14b of Paper II the backgrounds are spread out in a diagonal in an area where one would expect the age-extinction degeneracy to take place, while in Fig. \ref{fig:param_param_detections}b the overlap with this area is minor. \cite{2019ARA&A..57..227K} caution, that all observations are prone to selection biases in favor of more compact clusters, as well as a survival bias. Clusters of age $\sim$10 Myr might be only weakly bound, but these would be under-represented in cluster catalogs, because almost none would survive to reach ages more than a few tens of millions of years. Our CNN, however, detects numerous young extended objects.
\end{itemize}

\section{Conclusions}

We present a full pipeline of cluster search using a CNN, which is able to infer cluster's parameters (age, mass, extinctions, and size), as well as provide inference reliability estimates. The CNN is trained on mock star clusters superimposed on images of real backgrounds and is applied to multi-band images as a sliding window, without using precise photometric flux calibrations. This produces object candidate detection certainty maps, which are then filtered to obtain any number of cluster candidates, depending on recall (completeness) tolerance, estimated by artificial cluster tests.

We applied the star cluster detection pipeline to collect 3,380 cluster candidates in HST WFC3 observations of M83 in F336W, F438W, and F814W passbands, and used this data to analyze the spatial distributions of cluster parameters (age, mass, extinction, size) w.r.t. the galaxy's spiral arms and its center.

We have shown that an age gradient w.r.t. the spiral arms can be observed inside the corotation radius of the galaxy. The younger population of cluster candidates peaks at $\sim$0.4 kpc leading the spiral arms, while the older population is shifted towards $\gtrsim$0.7 kpc, this finding being consistent with the density wave theory of spiral arms. We also find a greater proportion of high extinction clusters on the dust lanes of the spiral arms, as well as more dense older clusters towards the galactic center.

The CNN-based approach naturally joins various techniques used in the study of star clusters -- detection, inference of evolutionary parameters, and size estimation -- into a single pipeline, which gives a coherent view of cluster populations.

\section*{Acknowledgements}

This research was funded by a grant (No. LAT-09/2016) from the Research Council of Lithuania. This research made use of Astropy, a community-developed core Python package for Astronomy \citep{astropy:2013, astropy:2018}. Some of the data presented in this paper were obtained from the Mikulski Archive for Space Telescopes (MAST). STScI is operated by the Association of Universities for Research in Astronomy, Inc., under NASA contract NAS5-26555. We are thankful to the anonymous referees who helped improve the paper.

\bibliography{library}{}
\bibliographystyle{aasjournal}

\end{document}